\documentclass[aps,floatfix,showpacs,nofootinbib,showkeys,superscriptaddress]{revtex4}
\usepackage[T1]{fontenc}
\usepackage{amssymb,amsfonts,amsmath,mathtext,enumerate,float,mathrsfs}
\usepackage[utf8]{inputenc}
\usepackage[english]{babel}
\usepackage{graphicx,color}
\usepackage{wrapfig}
\graphicspath{{./}}

\usepackage{epsfig}
\usepackage{epsf}
\usepackage{subfig} 
\usepackage{feynmp} 
\usepackage{pgf,pgfarrows,pgfnodes,pgfautomata,pgfheaps,multirow}
\usepackage{array}
\usepackage{bm}
\usepackage{latexsym}
\usepackage{pifont}

\newcommand{\be}{\begin{equation}}
\newcommand{\ee}{\end{equation}}
\newcommand{\bea}{\begin{eqnarray}}
\newcommand{\eea}{\end{eqnarray}}
\newcommand{\dd}{\mbox{d}} 
\newcommand\nn{\nonumber}

\begin{document}

\title{Processes of $e^+e^- \to (\eta, \eta', \eta(1295), \eta(1475))\gamma$ in the extended
Nambu -- Jona-Lasinio model}

\author{A.\ I.\ Ahmadov}
\email{ahmadov@theor.jinr.ru}
\affiliation{Bogoliubov Laboratory of Theoretical Physics, JINR, Dubna, 141980  Russia}
\affiliation{Institute of Physics, Azerbaijan National Academy of Science, Baku, Azerbaijan}

\author{D.\ G.\ Kostunin}
\email{dmitriy.kostunin@kit.edu, kostunin@theor.jinr.ru}
\affiliation{Institut f\"ur Kernphysik, Karlsruhe Institute of Technology (KIT), Germany}

\author{M.\ K.\ Volkov}
\email{volkov@theor.jinr.ru}
\affiliation{Bogoliubov Laboratory of Theoretical Physics, JINR, Dubna, 141980  Russia}

\begin{abstract}
In the extended NJL model the total cross sections of the processes
$e^+e^- \to (\eta, \eta', \eta(1295), \\
\eta(1405))\gamma$ at energies up to 2 GeV are calculated.
The intermediate vector mesons  $\rho (770)$, $\omega(782)$, $\phi (1020)$,
$\rho' (1450)$, $\omega' (1420)$ and $\phi' (1680)$
are taken into account. The latter three mesons are treated as the first radial excited states.
They are incorporated into the NJL model by means of polynomial form factors.
The calculation results are in satisfactory agreement with the experimental data obtained
with the SND detector at the VEPP-2M, Novosibirsk.
The predictions given for the process $e^+e^- \to (\eta',\eta(1295), \eta(1405)) \gamma$
have not been experimentally tested yet.
\end{abstract}

\date{\today}



\maketitle

\section{Introduction}
\label{Introduction}
Recently in \cite{kuraev,arbuzov,volkov3,volkov4,ahmadov}, it has been shown that
in the framework of the extended chiral NJL model \cite{volkov6,volkov7,volkov8}
the $e^+e^- \to (\pi^0, \pi^{0'}(1300)); \,\,\,(\pi^0, \pi^{0'}(1300))\gamma; \,\,\,
\pi(\pi,\pi'(1300)); \,\,\,\pi^0\omega$ and $\pi^0\rho^0$
processes are satisfactorily described at energies up to 1.6~GeV.
The contributions from the intermediate vector mesons in both the ground and
radially excited states were taken into account.
Radially excited states of mesons were introduced in the standard NJL model
with the polynomial form factor of the second order in the transverse momentum of
quarks \cite{volkov6,volkov7,volkov8}.
In this case, the slope parameter of this form factor was defined by the
conditions that the quark condensate did not change after the introduction of
excited meson states in the standard NJL model.
In contrast with other phenomenological models used for description of low-energy interactions of
mesons, the extended NJL model does not need to introduce additional arbitrary parameters
for the above processes (see, for instance, \cite{dolinsky,nachasov}).
Up to now the processes with pion production, as well as $\rho, \omega$, and $\gamma$ in the
final state have been considered.
To do this, it was sufficient to use the $SU(2)\times SU(2)$ NJL model.
To consider the processes with production of $\eta$ and $\eta'$ mesons in the final state,
one has to use the extended $U(3) \times U(3)$ NJL model.
In \cite{volkov12}, the photoproduction of $\eta(550)$, $\eta'(950)$, $\eta(1295)$, $\eta(1405)$ in colliding electron-positron beams was studied.
Unlike the previous works, here we had to use the mixing matrix of the
four isoscalar states of $\eta$ mesons \cite{volkov20,Volkov:2000ry}.

In this work, we will continue investigation in this direction, namely, we will calculate the
total cross section of the processes $e^+e^- \to (\eta, \eta', \eta(1295), \eta(1405))\gamma$,
with intermediate states of vector mesons, in both the ground and radial-excited states in the energy
region up to 2~GeV.
Our results for the process $e^+e^- \to \eta\gamma$ are in satisfactory agreement with the experimental data obtained in \cite{achasov}.
The predictions for the processes containing $\eta'$, $\eta(1295)$, $\eta(1405)$ are given.
The predictions can be verified in the ongoing experiment at VEPP-2000 (Novosibirsk).

\section{Lagrangian of the extended NJL model}
\label{Lagrangian}
\begin{figure}[h]
\includegraphics[width=0.45\linewidth]{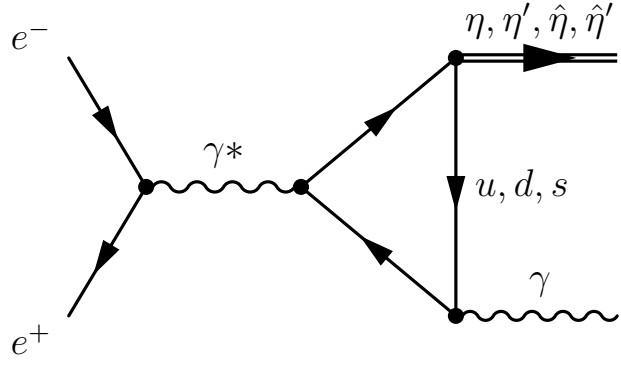}
\caption{Feynman diagram with photon exchange.}
\label{fig1}
\end{figure}

\begin{figure}[h]
\includegraphics[width=1.0\linewidth]{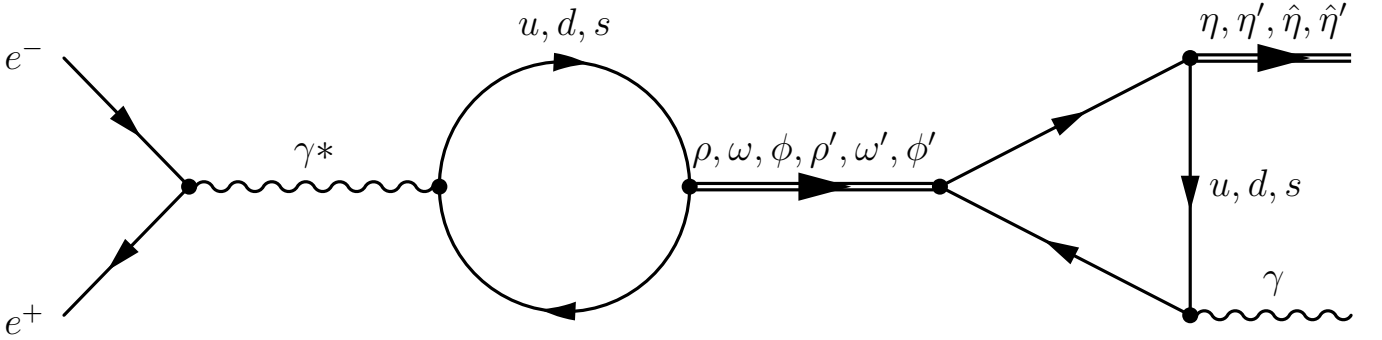}
\caption{Feynman diagrams with $\rho$, $\omega$, $\phi$, $\rho'$, $\omega'$, $\phi'$ meson exchange.}
\label{fig2}
\end{figure}

On Figs. \ref{fig1} and \ref{fig2} the processes $e^+ e^- \to (\eta, \eta', \hat\eta, \hat\eta')\gamma$ with taking into account
intermediate particles $\gamma$, $\rho$, $\omega$, $\phi$, $\rho'$, $\omega'$ and $\phi'$ are given.

For the description of these processes we will use the lagrangian of the extended $U(3)\times U(3)$ NJL model~\cite{volkov17,volkov19,volkov18,Volkov:2000ry}
\begin{eqnarray}
\Delta {\mathcal L}_2^{\mathrm{int}} &=& \bar{q}(k')\left( L_{\mathrm{f}} + L_\gamma + L_\mathrm{V} + L_{\eta,\eta',\hat\eta,\hat\eta'} \right) q(k) \,,\\
L_{\mathrm{f}} &=& i\hat\partial - m\,, \nn \\
L_\gamma &=& \frac{e}{2}\left(\lambda_3+\frac{\lambda_8}{\sqrt{3}}\right)\hat{A}\,, \nn \\
L_\mathrm{V} &=& A_{\omega,\rho}\left(\lambda_3{\hat{\rho}}(p)+\lambda_u\hat{\omega}(p)\right) + A_\phi \lambda_s\hat{\phi}(p) \nn \\ &-& A_{\omega',\rho'}\left(\lambda_3{\hat\rho'}(p)+\lambda_u\hat\omega'(p)\right) - A_{\phi'} \lambda_s\hat{\phi'}(p)\,, \nn \\
L_{\eta,\eta',\hat\eta,\hat\eta'} &=& i\gamma_5\sum\limits_{q=u,s}\lambda_q\sum\limits_{\boldsymbol\eta = \eta,\eta',\hat\eta,\hat\eta'}A_{\boldsymbol{\eta}}^q\eta(p), \nn
\end{eqnarray}
where $\bar{q}=(\bar{u},\bar{d}, \bar{s})$ with $u$, $d$ and $s$ quark fields;
$m=\mathrm{diag}(m_u,m_d,m_s)$, $m_u=m_d=280$~MeV and $m_s=455$~MeV are the constituent quark masses;
$e$ is the electron charge;
$A$ is the photon field, $\rho$, $\omega$, $\phi$, $\eta$, $\eta'$, $\hat\eta$, $\hat\eta'$ are meson fields, respectively (hat over $\eta$,$\eta'$ fields means an exited state);
\begin{eqnarray}
&& \lambda_3 = \left(\begin{array}{ccc}1& & \\ &-1& \\ & &0\end{array}\right), \lambda_8 = \frac{1}{\sqrt 3}\left(\begin{array}{ccc}1& & \\ &1& \\ & &-2\end{array}\right),\\
&& \lambda_u = \left(\begin{array}{ccc}1& & \\ &1& \\ & &0\end{array}\right),
\lambda_s = \left(\begin{array}{ccc}0& & \\ &0& \\ & &-\sqrt 2\end{array}\right), \nn
\end{eqnarray}
\begin{eqnarray}
A_{\omega,\rho} &=& g_{\rho_1}\frac{\sin(\beta^u+\beta_0^u)}{\sin(2\beta_0^u)}
       +g_{\rho_2}f_u({k^\bot}^2)\frac{\sin(\beta^u-\beta_0^u)}{\sin(2\beta_0^u)},
\\
A_{\omega',\rho'} &=& g_{\rho_1}\frac{\cos(\beta^u+\beta_0^u)}{\sin(2\beta_0^u)}
        +g_{\rho_2}f_u({k^\bot}^2)\frac{\cos(\beta^u-\beta_0^u)}{\sin(2\beta_0^u)},
\nn \\
A_{\phi} &=& g_{\phi_1}\frac{\sin(\beta^s+\beta_0^s)}{\sin(2\beta_0^s)}
       +g_{\phi_2}f_s({k^\bot}^2)\frac{\sin(\beta^s-\beta_0^s)}{\sin(2\beta_0^s)},
\nn \\
A_{\phi'} &=& g_{\phi_1}\frac{\cos(\beta^s+\beta_0^s)}{\sin(2\beta_0^s)}
        +g_{\phi_2}f_s({k^\bot}^2)\frac{\cos(\beta^s-\beta_0^s)}{\sin(2\beta_0^s)}, \nn
\end{eqnarray}
\begin{equation}
A_{\eta,\hat\eta,\eta',\hat\eta'}^q = g_{q_1}b_{\eta,\hat\eta,\eta',\hat\eta'}^{\varphi_{q,1}} + g_{q_2}b_{\eta,\hat\eta,\eta',\hat\eta'}^{\varphi_{q,2}}f_u({k^\bot}^2),
\end{equation}
where $q=u,s$.
The radially-excited states were represented in the NJL model using the form factor in the quark-meson interaction
\begin{eqnarray}
&& f_q({k^\bot}^2) = (1-d_q |{k^\bot}^2|) \Theta(\Lambda_3^2-|{k^\bot}^2|),
\\
&& {k^\bot} = k - \frac{(kp)p}{p^2},\,\, d_u = 1.788\ {\mathrm{GeV}}^{-2}, \,\, d_s = 1.727\ {\mathrm{GeV}}^{-2}, \nn
\end{eqnarray}
where $k$ and $p$ are the quark and meson momenta, respectively.
The cut-off parameter $\Lambda_3=1.03$~GeV.
The coupling constans are defined in the extended NJL model by the integrals containing given form-factors
\begin{eqnarray}
&& g_{q_1} = \left(4 \frac{I_2(m_q)}{Z_q}\right)^{-1/2},\quad g_{q_2} = \left(4 I_2^{f^2}(m_q)\right)^{-1/2}, \\
&& g_{\rho_1} = \left(\frac{2}{3} I_2(m_u)\right)^{-1/2},\quad g_{\rho_2} = \left(\frac{2}{3} I_2^{f^2}(m_u)\right)^{-1/2}, \nn \\
&& g_{\phi_1} = \left(\frac{2}{3} I_2(m_s)\right)^{-1/2},\quad g_{\phi_2} = \left(\frac{2}{3} I_2^{f^2}(m_s)\right)^{-1/2}, \nn
\end{eqnarray}
where $Z_q$ factor appeared after taking into account pseudoscalar -- axial-vector transitions, $Z_s \approx Z_u = 1.2$,
\begin{equation}
I^{f^{n}}_m(m_q) = -iN_c\int\frac{\dd^4 k}{(2\pi)^4}\frac{(f_q({k^\bot}^2))^n}{(m_q^2-k^2)^m}\Theta(\Lambda^2_3 - \vec k^2),
\end{equation}
where $N_c = 3$ is a number of quark colors. The mixing angles for vector mesons are $\beta_0^u = 61.44^{\circ}$, $\beta^u = 79.85^{\circ}$, $\beta_0^s = 57.11^{\circ}$, $\beta^s = 76.18^{\circ}$. One can find the definition of mixing angles for $\rho$, $\omega$ and $\phi$ mesons in ~\cite{volkov17,Volkov:2000ry}. The mixing coefficients for the isoscalar pseudoscalar meson states given in Table~\ref{table:1} were defined in~\cite{volkov18}.
\begin{table}
\caption{The mixing coefficients for the isoscalar pseudoscalar meson states.}
\begin{ruledtabular}
\begin{tabular}{rcccc}	 		
&$\eta$ 		&$\hat\eta$ 	&$\eta'$ 		&$\hat\eta'$\\
$\varphi_{u,1}$	&$0.71$		&$0.62$		&$-0.32$		&$0.56$		\\
$\varphi_{u,2}$   &$0.11$		&$-0.87$		&$-0.48$		&$-0.54$		\\
$\varphi_{s,1}$	&$0.62$		&$0.19$		&$0.56$		&$-0.67$		\\
$\varphi_{s,2}$   &$0.06$		&$-0.66$		&$0.30$		&$0.82$		\\
\end{tabular}
\end{ruledtabular}
\label{table:1}
\end{table}

It should be noted that meson states $\phi'(1680)$ are taken into account only for qualitative predictions
because their mass is very close to the masses of $\rho''(1690)$ and $\omega''(1670)$, which are not
described in the framework of our version of the extended NJL model.

\section{Amplitudes and cross-sections of the processes}
All amplitudes of the given processes have the form
\begin{equation}
T^{\lambda}=\bar {e}\gamma^{\mu}e \cdot  \frac{p_{\eta}^{\alpha}p_{\gamma}^{\beta}}
{m s}\cdot \{T_{\gamma}+T_{\rho + \omega} + T_{\phi}+T_{\rho'+\omega'}+T_{\phi'}\}\varepsilon_{\mu\lambda\alpha\beta},
\end{equation}
where $s=(p_+(e^+)+p_-(e^-))^2$.
The constributions from the photon and vector mesons read:
\begin{eqnarray}
T_\gamma &=& \frac{2}{3}\left (5 \frac{16}{3}\pi^2 m_u V_{\gamma u} + \sqrt{2}\frac{16}{3}\pi^2 m_s V_{\gamma s} \right),\\
T_{\rho + \omega} &=& \left(\frac{3s}{m_{\rho}^2 - s - i\sqrt{s}\Gamma_\rho} + \frac{1}{3}\frac{s}{m_{\omega}^2 - s - i\sqrt{s}\Gamma_\omega}\right) \nn \\  &\cdot&  \frac{C_{\gamma\rho}}{g_{\rho_1}} \left( \frac{16}{3}\pi^2 m_u V_{\rho} \right), \nn \\
T_{\phi} &=& -\frac{2\sqrt 2}{3}\frac{s}{m_{\phi}^2  - s - i\sqrt{s}\Gamma_\phi} \frac{C_{\gamma\phi}}{g_{\phi_1}} \left( \frac{16}{3}\pi^2   m_s V_{\phi}\right), \nn \\
T_{\rho' + \omega'} &=& \left(\frac{3s}{m_{\rho'}^2 - s - i\sqrt{s}\Gamma_{\rho'}(s)} + \frac{1}{3}\frac{s}{m_{\omega'}^2 - s  -i\sqrt{s}\Gamma_{\omega'}}\right) \nn \\ &\cdot& \frac{C_{\gamma\rho'}}{g_{\rho_1}} \left( \frac{16}{3}\pi^2 m_u V_{\rho'} \right) e^{i\pi}, \nn \\
T_{\phi'} &=& -\frac{2\sqrt 2}{3}\frac{s}{m_{\phi'}^2 - s - i\sqrt{s}\Gamma_{\phi'}}\frac{C_{\gamma\phi'}}{g_{\phi_1}} \left ( \frac{16}{3}\pi^2 m_s V_{\phi'}\right), \nn
\end{eqnarray}
where the coefficients $C_{\gamma V}$ denote the transitions of a photon into vector mesons
\begin{eqnarray}
&& C_{\gamma V_q} = \frac{\sin(\beta^{q}+\beta_0^{q})}{\sin(2\beta_0^{q})} + \Gamma_q\frac{\sin(\beta^{q}-\beta_0^{q})}{\sin(2\beta_0^{q})}, \\
&& C_{\gamma V'_q} = -\left(\frac{\cos(\beta^{q}+\beta_0^{q})}{\sin(2\beta_0^{q})} + \Gamma_q\frac{\cos(\beta^{q}-\beta_0^{q})}{\sin(2\beta_0^{q})}\right), \nn \\
&& \Gamma_q = \frac{I^{f}_2(m_q)}{\sqrt{I_2(m_q) I_2^{f^2}(m_q)}},\quad \Gamma_u = 0.54, \quad \Gamma_s = 0.41, \nn \\
&& q = u,s,\qquad V_u = \rho,\quad V'_u = \rho',\quad V_s = \phi,\quad V'_s = \phi'. \nn
\end{eqnarray}
The values of the triangles $V_{\gamma,\rho,\phi,\rho',\phi'} = V_{\gamma,\rho,\phi,\rho',\phi'}^{\eta,\eta',\hat\eta, \hat\eta'}$ depend on the outgoing particles. We give their values
\begin{eqnarray}
V_{\gamma q}^{\eta,\eta',\hat\eta, \hat\eta'} &=& \sum\limits_{i=1,2} b_{\eta,\hat\eta,\eta',\hat\eta'}^{\varphi_{q,i}} g_{q_i} I_3(m_q), \\
V_{V_q}^{\eta,\eta',\hat\eta, \hat\eta'} &=& \frac{\sin(\beta^{q}+\beta_0^{q})}{\sin(2\beta_0^{q})}b_{\eta,\hat\eta,\eta',\hat\eta'}^{\varphi_{q,1}}g_{V_1}g_{q_1} I_3(m_q) \\
&+& \frac{\sin(\beta^{q}-\beta_0^{q})}{\sin(2\beta_0^{q})}b_{\eta,\hat\eta,\eta',\hat\eta'}^{\varphi_{q,1}}g_{V_2}g_{q_1} I^{f}_3(m_q) \nn \\
&+& \frac{\sin(\beta^{q}+\beta_0^{q})}{\sin(2\beta_0^{q})}b_{\eta,\hat\eta,\eta',\hat\eta'}^{\varphi_{q,2}}g_{V_1}g_{q_2} I^{f}_3(m_q) \nn \\
&+& \frac{\sin(\beta^{q}-\beta_0^{q})}{\sin(2\beta_0^{q})}b_{\eta,\hat\eta,\eta',\hat\eta'}^{\varphi_{q,2}}g_{V_2}g_{q_2} I^{f^2}_3(m_q), \nn \\
-V_{V'_q}^{\eta,\eta',\hat\eta, \hat\eta'} &=&  \frac{\cos(\beta^{q}+\beta_0^{q})}{\sin(2\beta_0^{q})}b_{\eta,\hat\eta,\eta',\hat\eta'}^{\varphi_{q,1}}g_{V_1}g_{q_1} I_3(m_q) \\
&+& \frac{\cos(\beta^{q}-\beta_0^{q})}{\sin(2\beta_0^{q})}b_{\eta,\hat\eta,\eta',\hat\eta'}^{\varphi_{q,1}}g_{V_2}g_{q_1} I^{f}_3(m_q) \nn \\
&+& \frac{\cos(\beta^{q}+\beta_0^{q})}{\sin(2\beta_0^{q})}b_{\eta,\hat\eta,\eta',\hat\eta'}^{\varphi_{q,2}}g_{V_1}g_{q_2} I^{f}_3(m_q) \nn \\
&+& \frac{\cos(\beta^{q}-\beta_0^{q})}{\sin(2\beta_0^{q})}b_{\eta,\hat\eta,\eta',\hat\eta'}^{\varphi_{q,2}}g_{V_2}g_{q_2} I^{f^2}_3(m_q). \nn
\end{eqnarray}
For the $\rho'$ meson we use the width $\Gamma_{\rho'}(s)$ similar to~\cite{volkov4}
\begin{widetext}
\begin{eqnarray}
\Gamma_{\rho'}(s) &=& \Theta(2m_\pi - \sqrt{s}) \Gamma_{\rho'\to2\pi} \nn
\\ &+& \Theta(\sqrt{s} - 2m_\pi)(\Gamma_{\rho'\to 2\pi} + \Gamma_{\rho'\to\omega\pi}\frac{\sqrt{s}-2m_\pi}{m_\omega-m_\pi})\Theta(m_\omega + m_\pi - \sqrt{s})
\nn
\\ &+& \Theta(m_{\rho'} - \sqrt{s})\Theta(\sqrt{s} - m_\omega - m_\pi)
 \left(\Gamma_{\rho'\to 2\pi} + \Gamma_{\rho'\to\omega\pi}
 + (\Gamma_{\rho'} - \Gamma_{\rho'\to2\pi}  -\Gamma_{\rho'\to\omega\pi})
\nn
\frac{\sqrt{s}- m_\omega - m_\pi}{m_{\rho'} - m_\omega - m_\pi}\right)
\nn
\\ &+& \Theta(\sqrt{s} - m_{\rho'})\Gamma_{\rho'},
\end{eqnarray}
\end{widetext}
where $\Gamma_{\rho'\to 2\pi} = 22$~MeV and $\Gamma_{\rho'\to\omega\pi} = 75$ MeV were calculated in~\cite{volkov17},
we use the complete width $\Gamma_{\rho'} = 400$~MeV \cite{pdg}.
For the widths of $\omega'$ and $\phi'$ mesons  in Breit-Wigner formulae
we will use the more simple expression for the width of their decays.
Namely we will use $\omega'$ and $\phi'$ decays width value at
$\sqrt{s} = m_{\omega', \phi'}$.
This is justified by the fact that the contribution of the amplitude of the
intermediate $\omega'$ mesons order of magnitude  lower that the contribution
from the amplitude of the $\rho'$ in the intermediate state.
As for $\phi'$ meson, its contribution is significant only in the region
$\sqrt {s} > 1.5$~GeV.
For other mesons we use static widths. For all masses and decay widths we used the standard PDG values~\cite{pdg}. Let us note that the $\rho'$ and $\omega'$ mesons have simular masses and decay widths and $T_{\omega'} \approx 0.1 T_{\rho'}$. That means, the contribution from the $\omega'$ meson to the cross-section is negligible in present the processes. Unfortunatly, our model cannot describe phases of the excited intermediate resonances, thus we take the phases for $\rho'$ and $\omega'$ from the experiments~\cite{achasov,volkov3} $T_{\rho',\omega'}\to e^{i\pi}T_{\rho',\omega'}$.

\begin{figure}[h]
\begin{center}
\includegraphics[width=1.0\linewidth]{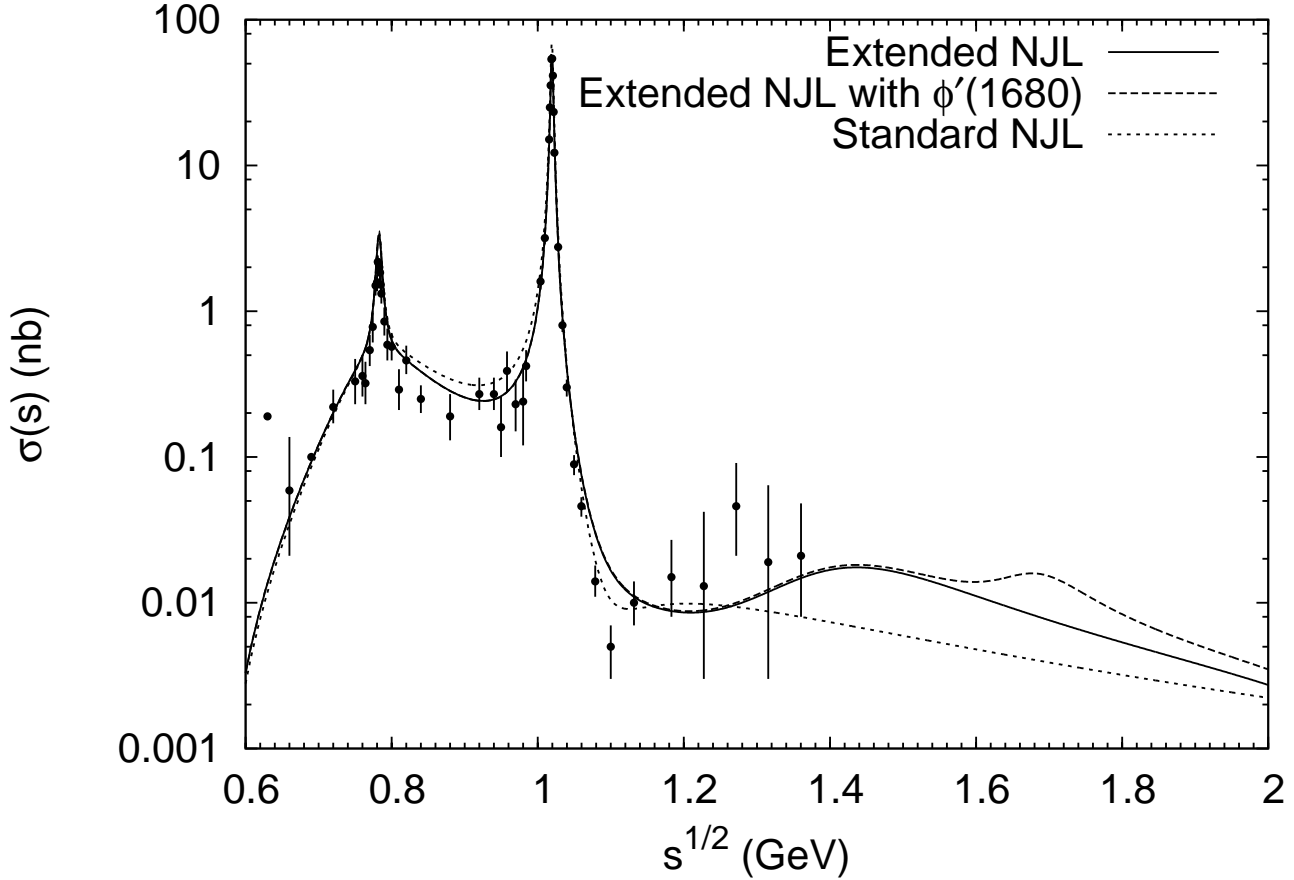}
\end{center}
\caption{Comparison of the NJL model predictions with the experiment~\cite{achasov} for the $e^+e^-\to\eta\gamma$ process.}
\label{plot1}
\end{figure}

\begin{figure}[h]
\begin{center}
\includegraphics[width=1.0\linewidth]{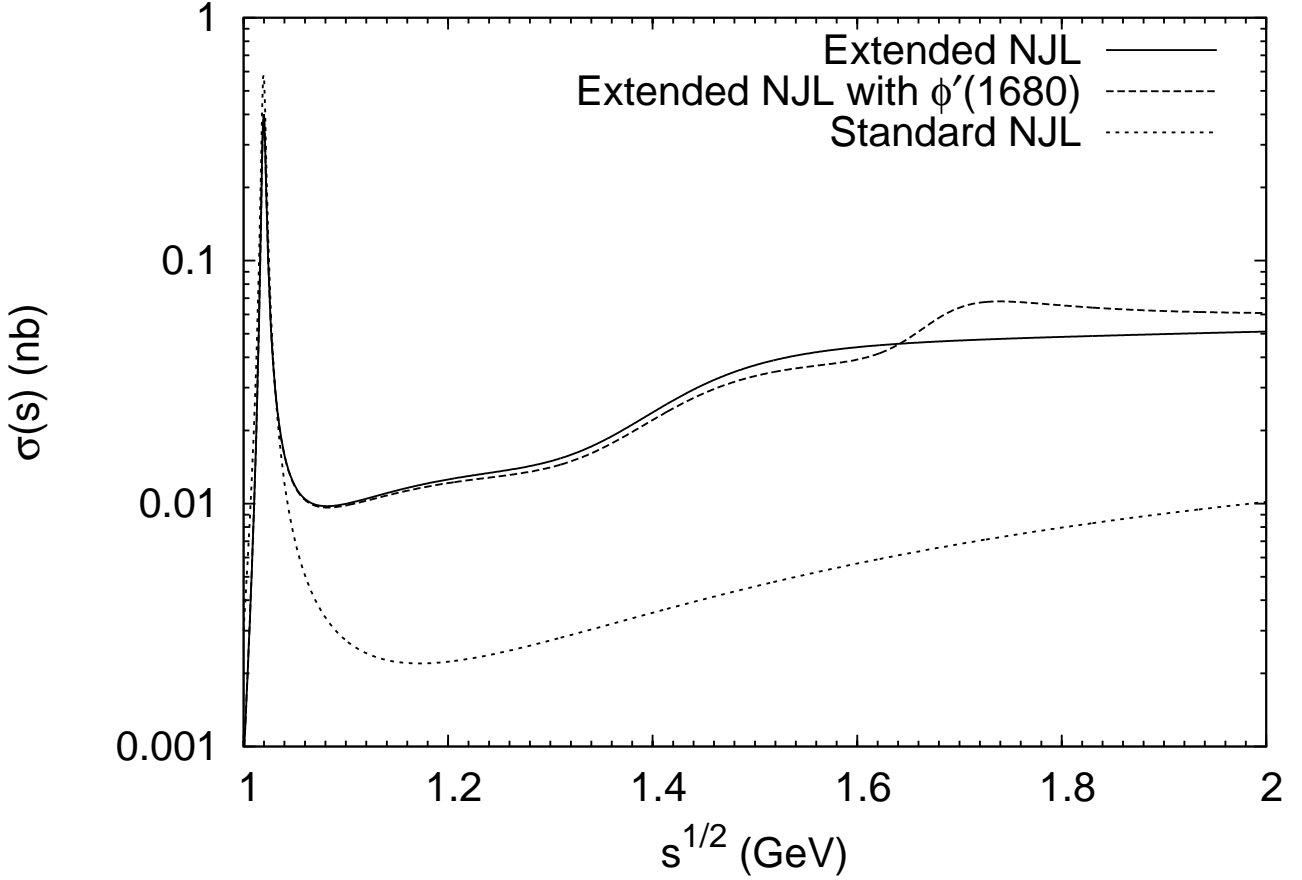}
\end{center}
\caption{Predictions for the $e^+e^-\to\eta'\gamma$ process given by the extended and standard NJL models.}
\label{plot2}
\end{figure}

\begin{figure}[h]
\begin{center}
\includegraphics[width=1.0\linewidth]{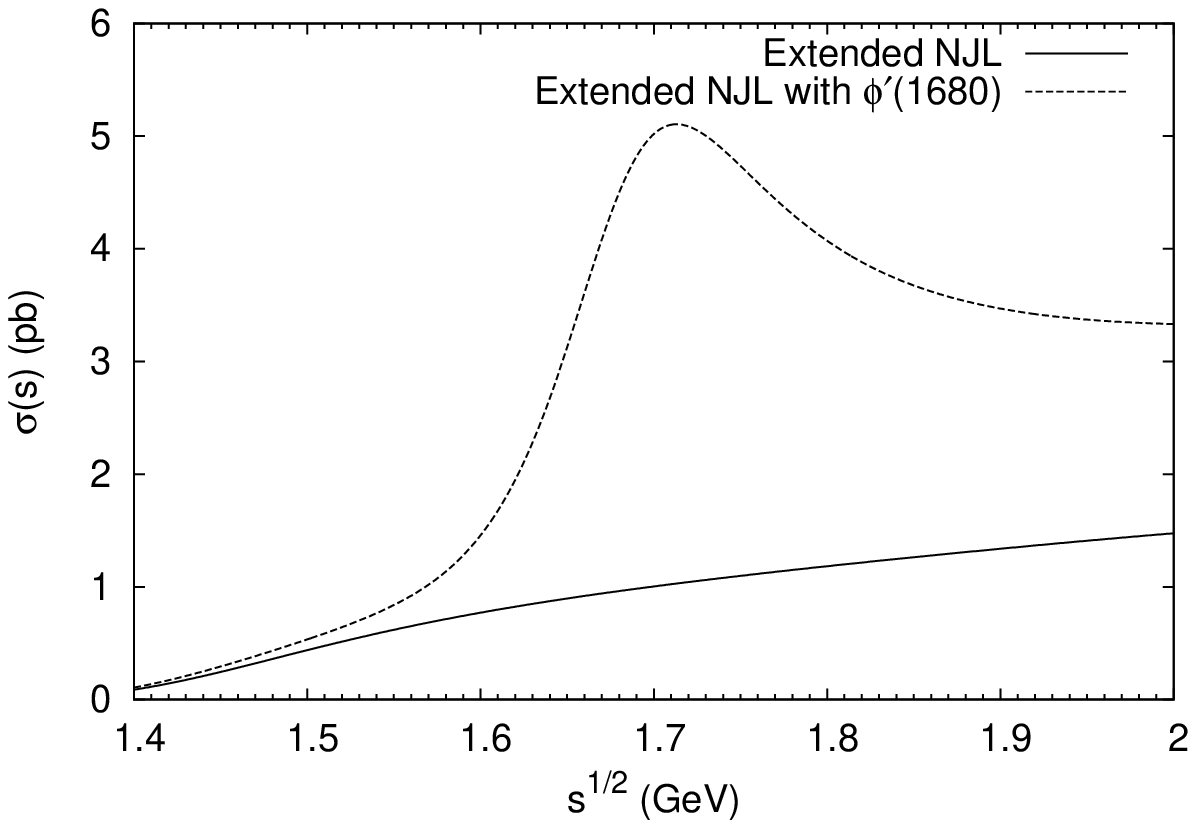}
\end{center}
\caption{Predictions for the $e^+e^-\to\eta(1295)\gamma$ process given by the extended NJL model.}
\label{plot3}
\end{figure}

\begin{figure}[h]
\begin{center}
\includegraphics[width=1.0\linewidth]{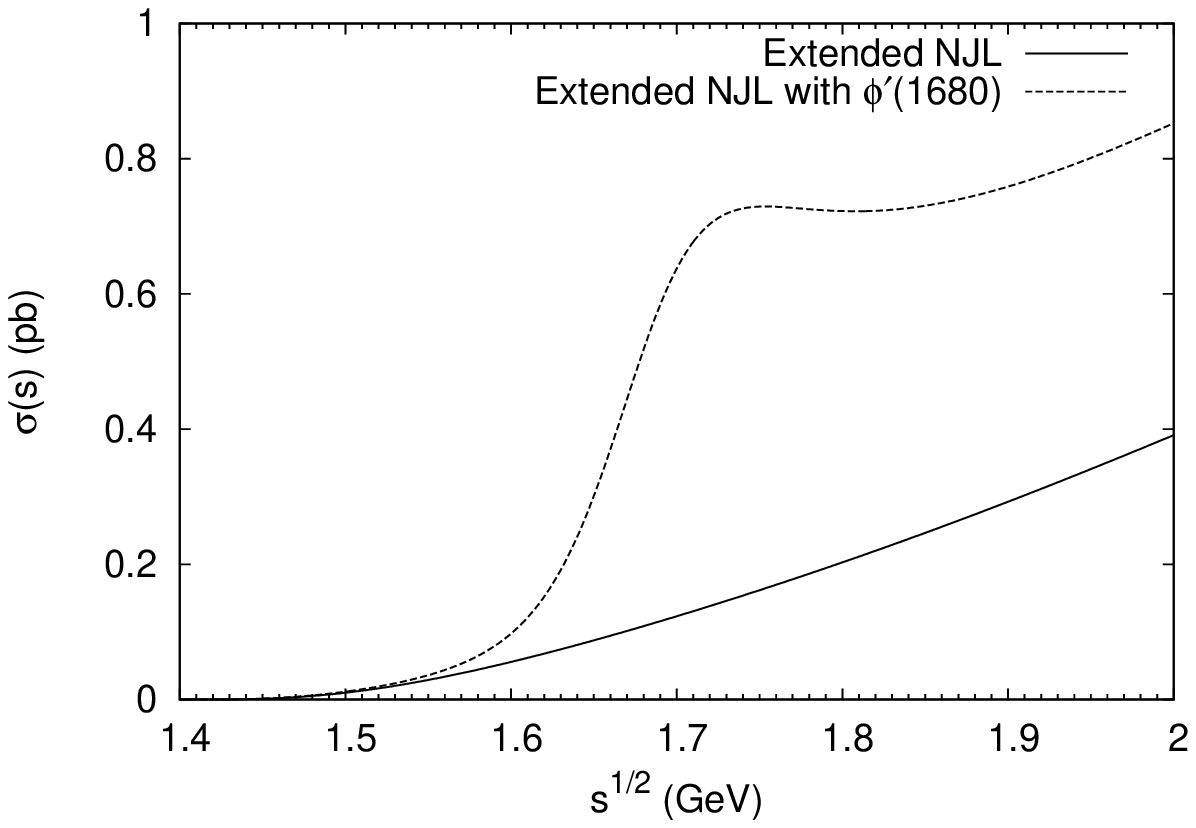}
\end{center}
\caption{Predictions for the $e^+e^-\to\eta(1405)\gamma$ process given by the extended NJL model.}
\label{plot4}
\end{figure}

The total cross-section takes the form
\begin{equation}
\sigma(s) = \frac{\alpha}{24\pi^2 s^3} \lambda^{3/2}(s,m,0)|T|^2,
\end{equation}
where $\lambda(a,b,c) = (a-b-c)^2 - 4bc$, $m = m_\eta, m_{\eta'}, m_{\hat\eta}, m_{\hat\eta'}$. One can find the cross-sections of each of four processes in Figs.~\ref{plot1},~\ref{plot2},~\ref{plot3},~\ref{plot4}. We calculated every cross-section up to 2~GeV, and give a comparison with experiment in Fig.~\ref{plot1} (experimental points shown in Fig.~\ref{plot1} are taken \cite{achasov}) and three
predictions for each process involving $\eta'$, $\eta(1295)$ and $\eta(1405)$ mesons (see Figs.~\ref{plot2},~\ref{plot3},~\ref{plot4}).
Every process was calculated with several different amplitudes. First amplitude was calculated in the extended NJL model but without the contribution of the $\phi'$ meson (solid curves). The second amplitude is obtained from the same amplitudes with taking into account contribution $T_{\phi'}$ from $\phi'$ meson (dashed curves). The third amplitude (in Figs.~\ref{plot1} and~\ref{plot2}) was obtained in the framework of the standard NJL (dotted curves). Note that in Fig.~\ref{plot1} one can see two sharp peaks. The first peak corresponds to the $\rho$ and $\omega$ intermediate resonant
contributions, the second one is due to the $\phi$ meson.

One can see that heavier resonances like $\phi'(1680)$ may play a significant role after 1.5~GeV, especially in processes with outgoing particles heavier than 1.2 GeV (see Figs.~\ref{plot3} and~\ref{plot4}).

\section{Conclusion}
The calculations resulted in satisfactory agreement with the experimental data relating to the measurement of the $e^+e^- \to \eta\gamma$ process in the energy region in the interval $0.6$ -- $1.4$ GeV.
The contributions of the intermediate vector mesons $\rho, \omega, \phi, \rho', \omega'$ and $\phi'$  were taken into account.
For the calculations of processes involving ground and exited states of the intermediate vector resonances we used the extended NJL model.

We demonstrate that the contributions of the amplitudes with the intermediate $\rho$, $\omega$ and $\phi$ mesons in the ground states give results being very close to the ones received within the extended model up to about
1.5 ~GeV, where the $\rho'$ resonance contribution becomes important.

It should be emphasized that our version of the extended model contains only the ground and the first radially excited states of the vector mesons.

We introduced radially excited $\phi'(1680)$ to show that heavier resonances like this one can make a notable contribution to cross-sections at larger energies.

In the considered here reactions, the main role is played by processes
with intermediate vector mesons. However, there are also reactions, where
intermediate scalar, tensor, and axial-vector mesons do contribute as
well, see e.g. Refs.~\cite{Oller:1997yg,Volkov:2012be,Ivanov:1989qw}.

We hope that our predictions for production finals states with $\eta'$,
$\eta(1295)$, and $\eta(1405)$ will be experimentally verified in
the exisiting and future experiments at $e^+ e^-$ colliders.


\section*{Appendix. Comparison of the results obtained in the standard and extended NJL model}
\label{appendix}	
In the appendix, we will show the calculation of the amplitude of the process $e^+e^-\to\eta\gamma$ in the framework of the standard NJL model. The lagrangian of the standard NJL model has the form~\cite{volkov9,volkov10,volkov11,volkov16,volkov18}
\begin{widetext}
\begin{eqnarray}
{\mathcal L}_{standard}^{\mathrm{int}} &=& \bar{q}\biggl[  i\hat\partial - m + \frac{e}{2}\left(\lambda_3+\frac{\lambda_8}{\sqrt{3}}\right)\hat{A} \\
&+& i\gamma_5\eta(\lambda_u g_\pi  \sin\bar\theta+\lambda_s g_s \cos\bar\theta) + i\gamma_5\eta'(\lambda_u g_\pi\cos\bar\theta - \lambda_s g_s\sin\bar\theta) \nn \\
&+& \frac{g_\rho}{2}(\lambda\hat\rho + \lambda_u\hat\omega) + \frac{g_\phi}{2}\lambda_s\hat\phi \biggr] q , \nn
\end{eqnarray}
\end{widetext}
where $\bar\theta = \theta_0 - \theta = 54^{\circ}$, $\theta$ is the singlet-octet mixing angle and $\theta_0$ is the ideal mixing angle; the coupling constants are $g_\pi = g_{u_1} = m_u/F_\pi$, $F_\pi=93$ MeV, $g_\rho = g_{\rho_1}$, $g_s = g_{s_1}$, $g_\phi = g_{\phi_1}$.

One can find the description of the transition of a photon into vector mesons into~\cite{volkov9}.
The amplitudes of the decays $\rho,\omega,\phi$ were calculated in~\cite{volkov10}.
We will follow the standard model and calculate convergent integrals which contain three quark propagators within the infinite limits
\begin{eqnarray}
I_3^{(\infty)}(m_q) &=& \frac{16}{3}\pi^2 m_q I_3(m_q)\biggr|_{\Lambda_3\to \infty} \\ \nn &=& \int\frac{\dd^4 k}{i\pi^2}\frac{m_q}{(m_q^2-k^2)^3} = \frac{1}{2m_q}.
\end{eqnarray}
Thus, the amplitude $T_{standard}$ for the given process contains the following terms:
\begin{eqnarray}
T_\gamma &=& \frac{1}{3}\left (5 \frac{g_\pi\sin\bar\theta}{m_u} + \sqrt{2}\frac{g_s\cos\bar\theta}{m_s} \right),\\
T_{\rho + \omega} &=& \frac{g_\pi}{2m_u}\left(\frac{3s}{m_{\rho}^2 -s - i\sqrt{s}\Gamma_\rho} + \frac{1}{3}\frac{s}{m_{\omega}^2 - s - i\sqrt{s}\Gamma_\omega}\right)\sin\bar\theta, \nn \\
T_{\phi} &=& \frac{\sqrt 2 g_s}{3m_s}\frac{s}{m_{\phi}^2-s - i\sqrt{s}\Gamma_\phi} \cos\bar\theta . \nn
\end{eqnarray}
It is worth noting that the amplitude with the intermediate photon (see Fig.~\ref{fig1}) in this case
can be easily combined with the amplitudes with the intermediate $\rho, \omega$, and  $\phi$ mesons \cite{volkov9}.
At the result, we reproduce the vector-dominance model picture:
\begin{widetext}
\begin{eqnarray}
&& T_{standard} = T_{\gamma} + T_{\rho + \omega} + T_{\phi} = T^{VMD}_{\rho + \omega} + T^{VMD}_{\phi}, \\
&& T_{\rho + \omega}^{VMD} = \frac{g_\pi}{2m_u}\left(3\frac{1 - i\sqrt{s}\Gamma_\rho/m_\rho^2}{m_{\rho}^2 -s - i\sqrt{s}\Gamma_\rho}m_\rho^2 + \frac{1}{3}\frac{1 - i\sqrt{s}\Gamma_\omega/m_\omega^2}{m_{\omega}^2 - s - i\sqrt{s}\Gamma_\omega}m_\omega^2\right)\sin\bar\theta, \nn \\
&& T_{\phi}^{VMD} = \frac{\sqrt 2 g_s}{3m_s}\frac{1 - i\sqrt{s}\Gamma_\phi/m_\phi^2}{m_{\phi}^2-s - i\sqrt{s}\Gamma_\phi}m_\phi^2 \cos\bar\theta . \nn
\end{eqnarray}
\end{widetext}
One can see the comparison between the predictions given by the extended and standard versions of the NJL model on Figs.~\ref{plot1} and~\ref{plot2}. Let us note that the standard NJL model gives us a value for the $\phi$-resonance peak larger than the experimental and predicted by the extended NJL model.

Using the formulas for the $V\to\eta\gamma$ vertices we can calculate the decay branching ratios for intermediate resonances into $\eta\gamma$ and compare them with experimental and PDG values. One can see a comparison in Table~\ref{tbl_comp}.
\begin{table}
\caption{The width decays for the processes $\rho\to\eta\gamma$, $\omega\to\eta\gamma$, $\phi\to\eta\gamma$ (in KeV).}
\begin{center}
\begin{ruledtabular}
\begin{tabular}{cccc}
Process & Standard NJL  & Extended NJL & PDG~\cite{pdg} \\
 $\Gamma(\rho\to\eta\gamma) $ & $72.9$ & $63.4 $ & $44.7 \pm 0.16$ \\
 $\Gamma(\omega\to\eta\gamma) $ & $ 8.7 $ & $7.6 $ & $ 3.9 \pm 0.032$ \\
 $\Gamma(\phi\to\eta\gamma) $ & $ 54.3 $ & $51.8 $ & $ 55.76 \pm 0.096$ \\
\end{tabular}
\end{ruledtabular}
\end{center}
\label{tbl_comp}
\end{table}

\section*{Acknowledgements}

We are grateful to A.~B.~Arbuzov, E.~A.~Kuraev, and V.~V.~Lenok for useful discussions.

\end{document}